# A True Assessment of Flat Lenses for Broadband Imaging Applications

JACOB ENGELBERG AND URIEL LEVY*

*Department of Applied Physics, The Center for Nanoscience and Nanotechnology, The Hebrew University, Jerusalem 91904, Israel*
*ulevy@mail.huji.ac.il*

**Abstract:** A plethora of metalenses and diffractive lenses ('flat lenses') have been demonstrated over the years. Recently, attempts have been made to stretch their performance envelope, particularly in the direction of wide-band achromatic performance. While achromatic behavior has been demonstrated, an actual improvement in imaging performance relative to conventional (non-chromatically corrected) flat lenses has not. The reasons for this are use of inappropriate performance metrics, lack of comparison to a baseline conventional design, and lack of a performance metric that combines signal-to-noise ratio and resolution. In this work we present a metric that will allow comparison of different types of flat lenses, even if their first order optical parameters are not the same. We apply this metric to several published achromatic flat lens designs and compare them to the equivalent conventional flat lens, which we consider as the lower bound for achromatic flat lens performance. Use of this metric paves the way for future developments in the field of achromatic flat lenses, which will display proven progress.

## 1. Introduction

Flat lenses, which can be implemented as diffractive lenses or metalenses, hold a great promise for miniaturization and economical mass production of optical systems by replacement of conventional refractive lenses [1–3]. However, the resolution of conventional non-chromatically corrected flat lenses, which we will call 'conventional flat lenses' (CFLs), is limited by their strong chromatic aberration [4]. This drawback motivates recent research on the development of achromatic flat lenses (AFLs) [5–7]. Unfortunately, the achromatization usually comes at the expense of reduced efficiency, lens power, and field-of-view (FOV).

Implementing an AFL, be it either an achromatic diffractive lens (ADL) or an achromatic metalens (AML) is challenging. Nevertheless, there have been several demonstrations of AMLs of different types such as dispersion engineered, spatial multiplexed, and extended-depth-of-focus [3,7–11]. Recently there have also been demonstrations of ADLs, leveraging the capabilities of extended phase depth and advanced optimization techniques [6]. A sketch of conventional vs. achromatic flat lens behavior, in terms of ray optics, is shown in Fig. 1 [12].

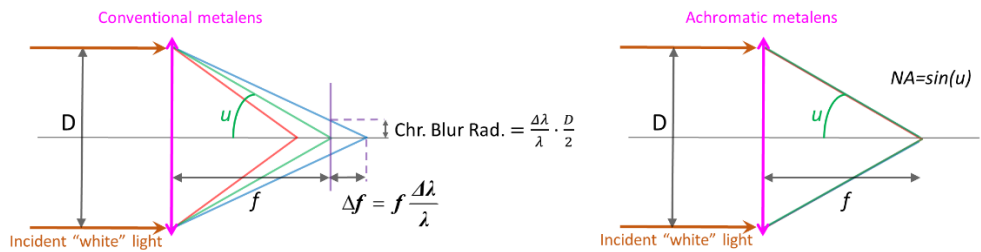

Fig. 1. Sketch showing ray paths, basic parameters (lens diameter D, focal length *f*, and numerical aperture NA) and chromatic aberration (longitudinal chromatic aberration *Δf*, and chromatic blur radius) for (a) conventional vs. (b) achromatic, diffractive lens.

The demonstrated AFLs have various first-order parameters, shown in Fig. 1: Aperture diameter $D$, focal length $f$, numerical aperture $NA = \sin(u) = \sin[\operatorname{atan}(D/2f)]$, central operating-wavelength (CWL) $\lambda_0$, spectral range $\Delta\lambda$, and sometimes a certain field-of-view (FOV, not drawn). The challenge in assessing the contribution of a new implementation or design lies in the difficulty to determine the level of progress beyond the state of the art (the demonstrated lenses are mostly generic and not intended for a specific application, so the innovation cannot usually be determined based on functionality). This causes research in the field of flat lenses to lack a clear path of progress.

As of today, the performance metrics used by most authors are the focusing efficiency and full-width half-maximum (FWHM) of a light spot in the image plane. The definitions of focusing efficiency and diffraction efficiency, and the relation between them, are discussed in section 1 of the Supplementary. In this paper, when we refer to efficiency, we mean classic diffraction efficiency, i.e., the fraction of the light incident on the lens aperture that goes to the design diffraction order. Efficiency represents the signal level but does not account for the background illumination, caused by spurious transmitted diffraction orders, that contribute to noise in an imaging system. Therefore, even if the efficiency is known, the effect of the flat lens on the signal-to-noise ratio (SNR) cannot be uniquely determined. Even worse, the FWHM of the measured spot (which is generally compared to the FWHM of the diffraction limited spot) which is supposed to represent the resolution, is in fact not an effective measure of resolution, as explained in section 2 of the Supplementary.

An additional problem with the methods used to evaluate flat lenses is that the signal/efficiency and resolution parameters are treated separately, although they are clearly coupled [13,14]. For a CFL used for imaging applications with broadband scene illumination, there is a tradeoff between signal and resolution, based on the spectral bandwidth used. A narrow bandwidth will provide good resolution (small chromatic aberration) at the price of low signal and SNR (since not many photons from the broadband scene illumination have frequencies within the limited bandwidth of the system), while a broad bandwidth will give high signal/SNR at the expense of low resolution due to aberrations. Such a low-resolution/high-SNR image can be converted to a high-resolution/low-SNR image by performing a deconvolution via digital image processing. A similar tradeoff exists for the relative aperture (i.e. the *NA* or the *F#*, defined as *F#=f/D*) of the CFL [12]. These tradeoffs are also relevant to AFLs since their achievable level of resolution and efficiency is related to the level of chromatic aberration before correction (which in turn depends on spectral range and aperture). More importantly for our analysis, AMLs generally achieve improved resolution at the expense of efficiency, since larger metalens nano-structure height and refractive index contrast increase the achievable resolution, but tend to decrease the efficiency [15]. To summarize, an overall performance metric that accounts for both resolution and SNR is needed for proper evaluation of flat lens performance. This is the main purpose of our manuscript.

Perhaps the greatest problem preventing understanding of the quality of achromatic designs is that they are seldom compared to the equivalent conventional design. A true comparison should, of course, use a metric that combines resolution and SNR. However, even at the more basic level of resolution only, a proper comparison is rarely made.

In most publications the quantitative evaluation of AFLs has been done at discrete wavelengths. While this may be appropriate for some non-imaging applications (such as relay optics in a WDM communication system), the most common application for achromatic lenses is imaging over a continuous spectrum. In this paper we focus on such applications, so our metric and examples are tailored to lenses operating over a continuous spectral range. However, the metric we shall present can be adapted to non-imaging applications as well.

Finally, it would be very useful to have a performance metric that will allow comparison not only of lenses with the same first order parameters, but also of lenses with different parameters. This will allow us to compare between published designs and assess the relative level of achievement, even if they do not have the same first-order parameters. As will be shown, by considering both SNR and resolution we can formulate such a metric.

Following recent research, in which the upper limits of AFL performance were explored [15,16], in this paper we compare such AFLs to the benchmark in the field, set by the equivalent CFL. Our purpose is to answer the following questions: Do current versions of AFLs perform better than CFLs for broadband imaging applications? To answer this question, we combine resolution and efficiency into a single metric that allows us to compare overall AFL performance to the benchmark of CFL performance. We also compare different AFL designs, to see which gives the best results. An important question that arises is how much room is left between the upper limit and the benchmark, i.e., can AFLs make a significant contribution relative to conventional flat lenses? This is discussed in [16], where it is found that not much room is indeed left. However, the upper bounds apply only to a certain class of AMLs and ADLs (single layer dispersion-engineered metalens and single surface multilevel diffractive lens respectively), so these limits could be surpassed by using other approaches. *The metric presented in this paper can allow us to evaluate the level of contribution of such future designs.*

In section 2 we introduce our overall performance metric (OPM), for comparison of AFLs with identical first order parameters. In section 3 we develop the extended overall performance metric (EOPM) for comparison of AFLs with different parameters. In section 4 we expand these metrics to color imaging applications, and to applications where the field-of-view is significant (imaging applications). In section 5 we apply these metrics to several published AFL designs and compare them to equivalent CFL designs. Analyses of additional published AFL designs for color imaging applications are presented in section 8 of the Supplementary. Finally, in section 6 we draw our conclusions.

## 2. Overall performance metric

In the following we describe an overall performance metric for imaging applications based on average signal-to-noise ratio (ASNR), as defined by Eq. 1 [17]. Here, the signal-to noise ratio (SNR) refers to zero (or low) spatial frequency SNR, where the modulation transfer function (MTF) takes a value of 1. If we produce an image of a large dark area on a white background, the noise is the standard deviation of the pixel gray-level values in the white image area, and the signal is the difference between the average white and black gray-level values. To obtain the SNR at higher spatial frequencies, we multiply this SNR by the MTF of the system, to account for the signal attenuation at these spatial frequencies. The noise, however, is assumed to be constant over all frequencies (we assume a shot noise limited system, which is a "white" noise, i.e., equal amplitude at all spatial frequencies). Therefore, the SNR at any given spatial frequency is equal to the zero-frequency SNR multiplied by the MTF at that frequency. To obtain the ASNR one needs to average over all the spatial frequencies, as shown in Eq. 1. This expression is similar to what is obtained from an information theory approach, only that here we omit the *log* function [18]. We will revisit this in section 4, where we discuss extended FOV systems.

$$ASNR = \int_0^{v_{nq}} SNR \cdot MTF(v) dv / v_{nq} \quad , \quad v_{nq} = \frac{1}{2 \cdot \text{pixel\_pitch}} \tag{1}$$

The zero-frequency SNR in Eq. 1 is frequency independent, so it can go outside the integral. The integral limits, going from zero to Nyquist frequency, $v_{nq}$, are relevant when working with a specific camera. However, if we want to make an evaluation of the metalens regardless of the camera being used, we can take this limit to be the lens MTF cutoff (i.e., we assume the camera has small enough pixels such that the metalens rather the camera is limiting the resolution). In this case the area under the lens MTF, relative to the area under the diffraction limited MTF, is equal to the one-dimensional version of the Strehl ratio – see section 2 of Supplementary. Since the diffraction limited MTF is determined by the first-order lens parameters (NA and spectrum), we can use the Strehl ratio as a single number measure of resolution, instead of the MTF, if we compare lenses with the same parameters.

The SNR in the shot-noise limited case is proportional to the square root of the number of photons reaching a single camera pixel [12]. The absolute number of photons is dependent on many variables, but if we compare two flat lens systems with the same ambient and system parameters (target illumination, NA, spectral range, camera pixel size and integration time) we can say that the number of photons is proportional only to the flat lens efficiency [19].

The last statement is true if the incident light that does not reach the desired focal point is reflected or absorbed. However, in most realistic imaging applications a significant amount of light is lost to transmitted spurious diffraction orders, causing veiling glare (VG) [20,21]. We denote the total transmitted fraction of the incident light (going to all transmitted diffraction orders) as $T$, and the fraction of incident light that goes to the desired order of diffraction (usually the first order) as $\eta$. The SNR is then proportional to $\eta/\sqrt{T}$, since the total number of photons contributing to noise is proportional to $T$, while the signal is proportional to $\eta$ (Alternatively, we can reach the same expression by saying that the overall SNR is proportional to $\sqrt{T}$, when we initially consider all the light as signal. When we then take into account that our signal is only part of the total transmitted light, usually the first order of diffraction, the spurious diffraction orders reduce the contrast, i.e. the un-normalized MTF, by a factor of $\eta/T$ [20]. When multiplied, this gives the same $\eta/\sqrt{T}$ factor). See section 3 of Supplementary for discussion of how to measure this factor.

Based on the above considerations, we can define an overall performance metric (OPM) that is proportional to the ASNR according to Eq. 2.

$$OPM = \frac{\eta}{\sqrt{T}} \cdot Strehl \qquad (2)$$

Since in most publications no information is given regarding the overall transmission $T$, we are forced to take the optimistic assumption that there is no VG, i.e., $T = \eta$, in our analyses of published results shown in the following sections. In such a case, one obtains $OPM = \sqrt{\eta} \cdot Strehl$. However, it must be remembered that these are optimistic results. The other extreme case would be to assume $T = 1$. In such a case the difference between transmission and efficiency is attributed to VG, and the OPM is given by $OPM = \eta \cdot Strehl$ which is always smaller than the previous expression.

The OPM equals 1 in an ideal case, and 0 in the worst case. The values for $\eta$ and $T$ used in Eq. 2 should be a weighted average over the spectral range, based on the detector spectral responsivity. It is important to note that the above merit function is only effective for comparing metalenses of different types (such as CDL vs. ADL/AML) that share identical first-order parameters (relative aperture, focal length, spectral range, field-of-view). If one wants to quantitatively compare lenses of different parameters, the extended metric presented next should be used.

## 3. Extended overall performance metric

To devise a metric that can be used to compare the performance of lenses with different first order parameters, we use the following guidelines: (a) The signal of a scene illuminated by a broad spectrum is proportional to the spectral range $\Delta\lambda$ that is allowed to pass through to the detector [12]. We elect to introduce a relative spectral range factor of $\sqrt{\Delta\lambda/\lambda}$ into our metric, thus keeping the merit function unitless, but limiting it to comparison of lenses operating around the same CWL (the radical is because the SNR is proportional to the square root of the number of photons). (b) The larger the relative aperture the higher the SNR. The signal is proportional to the $NA^2$ [22,23], so the SNR will be proportional to the NA. (c) As stated in the previous section, use of the Strehl ratio instead of the area under the MTF is only justified if we are comparing lenses of the same parameters. If we are comparing lenses with different parameters, we must multiply the Strehl ratio by the area under diffraction limited MTF, to obtain the area under the aberrated MTF. We can approximate this area by using the diffraction limited MTF for a square aperture, which is a triangle having an area of $v_{co}/2 = NA/\lambda$, where $v_{co}$ is the diffraction limit MTF cutoff (the factor between the areas under MTF of round aperture to MTF of square aperture is $8/(3\pi)$) [23,24]. In effect, to compare lenses of different parameters, what we want is a merit function that is related not to the average SNR but rather to the integrated SNR (over spatial frequencies). Thus, unlike the expression for ASNR (Eq. 1), one should not divide by the frequency range. (d) The Strehl ratio and diffraction limit cutoff frequency only account for the image space resolution. However, what is really of interest is the object space resolution. For an imaging lens viewing a distant object the magnification is proportional to the lens focal length *f*, so we add this factor to our merit function (for other imaging applications, such as a microscope objective, the focal length alone does not determine the magnification, since it depends on the tube lens focal length. However, a short focal length generally means a small field-of-view and short working distance, so a larger *f* has an advantage there too).

Implementing the above-mentioned guidelines, we now arrive at eq. 3, which defines the extended OPM (EOPM) which allows us to compare between lenses having different first order parameters, limited only to the same CWL.

$$EOPM = f \cdot NA \cdot \sqrt{\frac{\Delta\lambda}{\lambda} \cdot \frac{NA}{\lambda}} \cdot OPM \qquad (3)$$

To simplify Eq. 3, we can use the Fresnel number (FN), given by Eq. 4 [25]. The FN is approximately equal to the maximum (unwrapped) phase induced in the wavefront by the lens, modulo $\pi$, so the higher the FN the greater the "work" the lens is doing [26]. Substituting Eq. 4 into Eq. 3 we obtain Eq. 5.

$$FN \approx \frac{f \cdot NA^2}{\lambda} \qquad (4)$$

$$EOPM = FN \sqrt{\frac{\Delta\lambda}{\lambda}} \cdot OPM \qquad (5)$$

As opposed to the OPM, which gives a value between 0 and 1, the EOPM is not limited by an upper bound. As previously mentioned, the EOPM is limited to comparing lenses with the same central operating wavelength (CWL). However, this is not a severe limitation, since it is generally not relevant to compare lenses operating at different spectral ranges, because of the different operating conditions (different scene illumination, available materials for optics and detector etc.). Note that it is not clear how $\Delta\lambda$ should be defined. While for a Gaussian distribution one may simply use the FWHM as the criterion of choice, this is not the case for a general spectral distribution. The more general choice is to use the standard deviation of the spectral response when viewed as a probability distribution, multiplied by a constant. The definition of the constant is flexible, as long it is used consistently, since the EOPM is relative. For the simulations presented in this paper, we chose the constant to be $2\sqrt{3}$, which is identical to the full width of a uniform ('Rect' function) spectral distribution.

Not surprisingly, the various parameters that go into the EOPM not only determine the performance, but also the design and manufacturing challenge. The larger the relative spectral range ($\Delta\lambda/\lambda$) the more challenging the achromatic design is, since the chromatic blur spot is proportional to this ratio [12]. The larger the relative aperture (high *NA* or lower *F#*) the more challenging the design, since the diffraction limited spot size gets smaller, and the geometrical aberrations become more significant. The larger the lens (i.e. the larger the absolute value of the focal length/aperture) the more challenging the design (for a specific NA), since geometrical aberrations scale with lens dimensions while the diffraction limit does not [27,28].

Note that the EOPM can be used not only to evaluate the performance of a design, but also to optimize its parameters. In a previous paper we showed how such optimization can be performed based on the ASNR metric [12]. By maximizing the EOPM we can perform a similar optimization, under the assumption that the resolution is limited by the optics rather than by the camera. For example, the maximum EOPM can give an indication of which aperture to choose given the spectral range, and vice versa. This method can be applied to both conventional and achromatic flat lenses. In Fig. 2 we present the EOPM calculated for a simple binary (two-phase levels) CDL, with a focal length of 5.2mm operating around a central wavelength of 550nm (the general shape of the graph is similar to the ASNR graphs shown in [12,17]). The absolute value of the EOPM will change for a multilevel CDL, since the efficiency will improve, but the shape of the graph will remain the same.

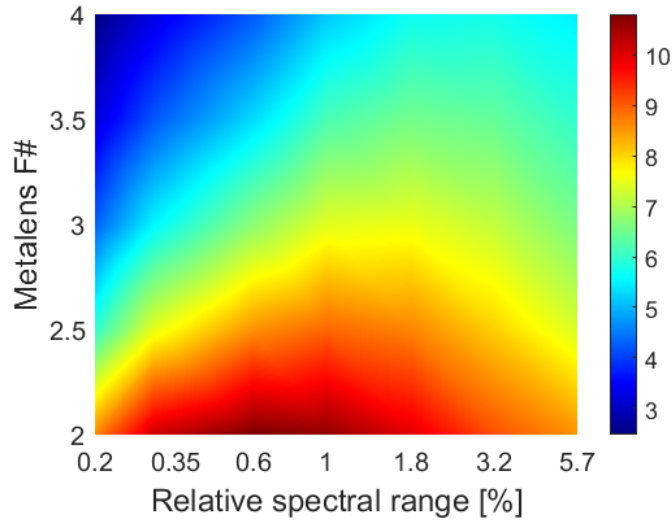

Fig. 2. EOPM for a CDL as a function of F# and relative spectral bandwidth $\Delta\lambda/\lambda$. For a given F#, one can determine the optimal operating waveband.

Based on Fig. 2, for a given F# and central wavelength, there is an optimal spectral bandwidth that should be used to maximize the CDL performance. However, it seems that the lower the F# the better the performance. What limits the F# on the low end? There are of course manufacturing/spatial sampling considerations, since for low F# one needs a very small diffractive period. However, even if we consider ideal manufacturing capabilities, there is a limit derived from field-of-view (FOV) considerations. The lower the F# the more difficult it becomes to obtain good performance over a reasonable FOV. Even for a system designed to operate on-axis only, one must account for this, because of tolerance issues. The way to do this is discussed in section 4.

As can be seen from Fig. 2, the EOPM varies even for a given flat lens technology. Although the EOPM can be used to compare lenses made using various technologies, it also includes the effects of the system parameters. Therefore, one cannot compare the EOPM for two lenses of identical parameters and conclude that a specific technology is always superior to the other. To compare two competing technologies, one must find the optimum EOPM for each of them, while allowing the first order lens parameters to vary over the available/practical range (e.g., if we want to make a camera lens with a specific FOV, we can vary the aperture and the wavelength range, as in Fig. 2). If relevant, the EOPM*fov* described in section 4 should be used. The technology that produces a better EOPM is the preferred one for the task, at least in terms of optical performance (practical considerations, such as manufacturing cost must be considered separately).

## 4. Expansions of performance metrics

In this paper we focus on comparing AFL and CFL performance for the case of broadband imaging applications with a small FOV. However, the OPM and EOPM metrics can be adapted and used for other applications as well. For example, for the case of a non-imaging system, such as a WDM fiber coupling lens, one may replace the 1D Strehl ratio with an overlap integral (for single-mode fiber), or with a parameter related to encircled energy (for multi-mode fiber). If the operation can be limited to discrete wavelengths, the performance should be evaluated and averaged over these wavelengths only.

The metrics can also be applied to systems operating quasi-monochromatically, where chromatic aberration is not an issue, and no added energy is gained by opening the spectral range. In this case, the $\sqrt{\Delta\lambda/\lambda}$ factor can be omitted.

A particularly interesting application is color imaging. For a color imaging system that has three separate channels, red (R), green (G) and blue (B), we can extend the above-mentioned discussion and use an average EOPM of the channels (with $\Delta\lambda/\lambda$ referring to the spectral range of each channel) to obtain the quality of the intensity component of the image (equal to R+G+B, where each letter represents the relevant gray-level image). However, when performing color imaging there is another performance parameter of interest: color fidelity. To obtain good color fidelity, it is not sufficient to have good average EOPM, rather one must have good EOPM for each individual channel. Therefore, for color imaging we suggest the metric of Eq. 6, which not only takes into account the average EOPM but also the standard deviation of the three channels, in such a way that it decreases with the increase in the standard deviation of the EOPM. The correction term (in the square brackets) multiplying the EOPM was chosen such that it will be zero when the EOPM of two out of three channels is zero (meaning we have no color information), and 1 when all three EOPMs are equal (for derivation see section 4 of Supplementary). If one is comparing color imaging lenses of identical first order parameters, the EOPM in Eq. 6 can be replaced with the OPM.

$$EOPMcolor = avg(EOPM)\left[1 - \frac{std(EOPM)}{\sqrt{3}avg(EOPM)}\right] \qquad (6)$$

Another factor that affects color fidelity is the shape of the spectral response of the flat lens [29,30], but that is beyond the scope of this paper. Here we assume that the spectral response of the lens is sufficiently broad so that the overall response is dominated by the camera, which has assumedly been designed to provide good color fidelity. Of course, it is possible to formulate other performance metrics for a color system, for example by replacing the 1 in the brackets with a 2, thus allotting some value to a system that gives a black and white image and giving extra credit for displaying color.

An additional important application are systems with a significant FOV. While in this paper we will apply the merit functions on-axis only, they can be extended to consider the FOV. The question is how does one account for the benefit of a larger FOV system? Defining a merit function (*EPOMfov*) for an optical system with a field-of-view is a bit tricky, since as mentioned in section 2, it is now necessary to apply a *log* function to the integrand in Eq. 1, to be consistent with information theory. As explained by Hartley [31], the need for the *log* function arises from the fact that although the number of permutations possible in an image of $g$ gray-levels and $p$ pixels is equal to $g^p$, the objective amount of information increases only linearly with $p$. The information content is therefore better represented by $log(g^p) = p \cdot log(g)$. This is also in line with the logarithmic sensitivity of the human eye. Use of a base 2 logarithm is convenient since it represents the number of bits necessary to transmit the information, although for our purposes any base can be used.

We therefore suggest the *EOPMfov* figure of merit given by Eq. 7, where *A* is the image area, based on the development in section 5 of the Supplementary. This figure of merit is more limited that the on-axis EOPM, in that the camera parameters and scene illumination must be considered in calculating the SNR, see Eq. S11. Lenses must be compared over the same image area *A*, but this does not limit the generality. If the MTF varies significantly over the FOV, the FOV must be broken up into isoplanatic patches (areas over which the MTF is nearly constant). In such a case the figure of merit should be calculated for each area separately, and then summed [32]. Note that the focal length factor is not included in this merit function. It is no longer necessary since the image area fulfills a similar function (as mentioned in section 3, the advantage of large *f* is mainly that it allows larger FOV, which is accounted for by *A*).

$$EOPMfov = A \iint\limits_{-v_{co}}^{v_{co}} log[1 + SNR^2 \cdot MTF^2(v_x, v_y)] dv_x dv_y \qquad (7)$$

## 5. Analysis of published achromatic flat lens results

To demonstrate the utility of our proposed metric, we apply it to several published AFL designs. Each time we will compare the AFL to the baseline, i.e., the equivalent (same first order optical parameters) CDL. This will allow us to evaluate the contribution (if any) of the achromatic design to the improvement of the optical performance for the chosen parameters.

We begin with three AML designs for the short-wave infrared (SWIR) spectral range presented in [5]. The optical parameters for the designs are summarized in the top rows of Table 1. We included the Airy radius and the chromatic aberration radius in the table, since these can be calculated using simple formulas, and allow a quick check of the level of chromatic aberration relative to the diffraction limit. Obviously, if the chromatic aberration is not much larger than the Airy radius, there is less need to correct the chromatic aberration.

The published measured performance is summarized under the column titled 'AML' in the bottom rows of Table 1. This performance is compared to the simulated performance of a classic 8 phase-level CDL (It would have been more appropriate to use simulated performance data for the metalens, since we are comparing to simulation, but none were given in this publication. However, for our CDL designs it is reasonable to assume performance close to theory, since there exists a well-established manufacturing technology). The optical performance parameters (PSF, MTF and Strehl ratio) for the CDLs were calculated using commercial optical design software (see section 6 of Supplementary for details). The results are shown in Fig. 3, and summarized in Table 1, under the column titled 'CDL'. The efficiency of the CDL was calculated analytically based on Eq. 12 of [33] (accounting for Fresnel reflection, number of phase levels and the shadow effect) and the spectral range (see section 7 of Supplementary for details). Uniform spectral response over the wavelength range was assumed. In principle we could have compared to a conventional (non-chromatically corrected) metalens, rather than to a diffractive lens. The reason we chose the diffractive lens option is because the efficiency can be calculated analytically.

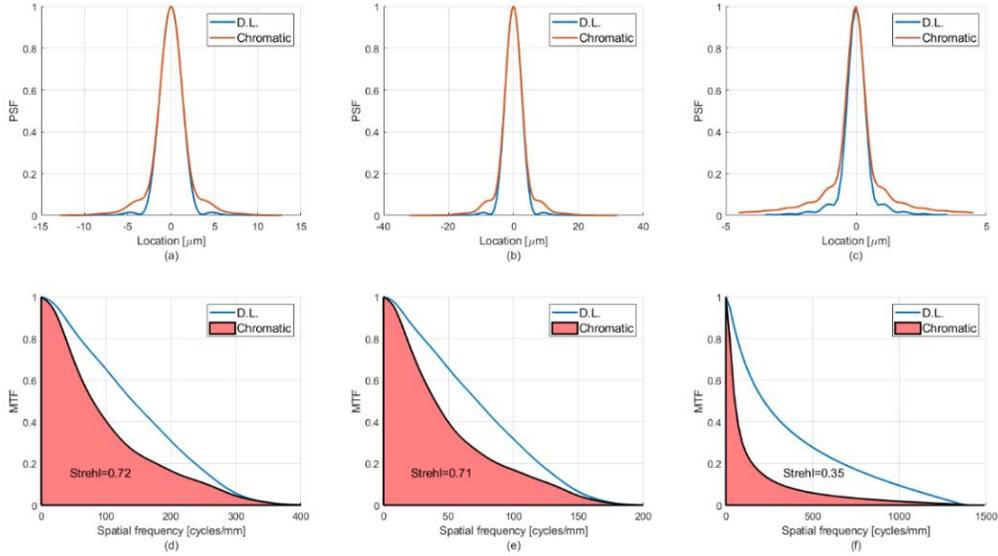

Fig. 3. PSFs, MTFs and 1D Strehl ratio of CDL designs equivalent to AML designs of [5], compared to the diffraction limit. (a) PSF cross section of CDL equivalent to AML design M1B compared to diffraction limited PSF (b) The same for design M2 (c) The same for design M3 (d) MTF of CDL equivalent to M1B compared to diffraction limited MTF, showing the 1D Strehl ratio, which is the ratio of shaded pink area to the area under the D.L. graph. (e) The same for design M2. (f) The same for design M3.

Table 1. Comparison of AMLs of [5] to equivalent CDLs

| Design | M1B | | M2 | | M3 | |
|---|---|---|---|---|---|---|
| λmin [µm] | 1.2 | | 1.2 | | 1.2 | |
| λmax [µm] | 1.65 | | 1.65 | | 1.4 | |
| NA | 0.24 | | 0.13 | | 0.88 | |
| EFL [µm] | 200 | | 800 | | 30 | |
| Dia. [µm] | 100 | | 210 | | 111 | |
| Fresnel no. | 8 | | 10 | | 51 | |
| Airy rad [µm] | 3.6 | | 6.7 | | 0.9 | |
| Chr. rad [µm] | 7.8 | | 16.6 | | 4.3 | |
| | AML | CDL | AML | CDL | AML | CDL |
| Efficiency | 0.35 | 0.78 | 0.35 | 0.87 | ? | 0.05 |
| 2D Strehl | 0.85 | 0.57 | 0.85 | 0.61 | ? | 0.16 |
| 1D Strehl | ~0.93 | 0.72 | ~0.93 | 0.71 | ? | 0.35 |
| OPM | 0.55 | 0.64 | 0.55 | 0.66 | | 0.08 |
| EOPM | 2.6 | 3.0 | 3 | 3.6 | | 1.5 |

It can be seen from Fig. 3 that for the first two designs (M1B and M2) the chromatic aberration of the equivalent CDL is not so large, so the resolution is not far from the diffraction limit, albeit not diffraction limited (the 2D Strehl is less than 0.8, which is generally taken as the threshold for diffraction limited performance). While the AML improves the resolution (higher Strehl ratio), the price paid in reduced efficiency makes the overall imaging performance worse than the CDL (lower OPM). In [5] the 2D Strehl ratio was given, and not the 1D. To compute our metrics, we estimated the 1D Strehl to be the average between the 2D Strehl and 1 (the 1D Strehl is almost always higher than the 2D, since the 2D Strehl overemphasizes the high frequencies, which naturally have lower MTF – see section 2 of

Supplementary). In our experience the average between the 2D Strehl and 1 gives an optimistic value for the 1D Strehl).

The third design (M3) is a high-NA design with more significant chromatic aberration, i.e., there is greater potential for improvement by using an AML. Unfortunately, the authors did not present resolution data for this design, claiming that Strehl ratio is not applicable to high-NA lenses. We beg to differ with this claim. If there is a measured or simulated PSF, and a diffraction limited PSF can be calculated, there is a Strehl ratio [24,34,35]. It is true that the calculation is more complicated in the high-NA case, as explained in section 6 of Supplementary, giving rise to the non-typical "belly" in the shape of the diffraction limited MTF (Fig. 3f and Fig. 4f). The only case when the Strehl ratio is not relevant is when the scalar approximation breaks down, i.e. the lens aperture is on the order of the wavelength, since in such a case the diffraction limited PSF cannot be calculated [36].

The efficiency of design M3 was also not measured, so we cannot compare to the CDL efficiency. The low CDL efficiency is caused primarily by poor phase sampling near the edge of the lens (we assumed a minimum feature size of 1µm) and includes the shadowing effect. Both effects can be partially overcome by a metalens, however, a metalens has other issues, such as those causing the low efficiencies in designs M1B and M2.

We now move on to a paper by Banerji et al. that compares the performance of metalens designs to equivalent diffractive lens designs [37]. Specifically, we apply our metric to the three broadband designs that were presented in [37]. It can be seen from Fig. 4 and Table 2 that for all three designs the CDL resolution is not far from the diffraction limit. The quoted AMLs give only a small improvement in resolution over the CDLs, and the price paid in efficiency is large, so the overall performance is worse. The ADLs designed by Banerji et al. improve upon the AMLs in terms of efficiency, but no real indication is given of the resolution performance, since the authors use the FWHM metric. The numbers marked in blue in the table show the 1D Strehl ratio that would be needed for the ADL designs to exceed the performance of the CDLs.

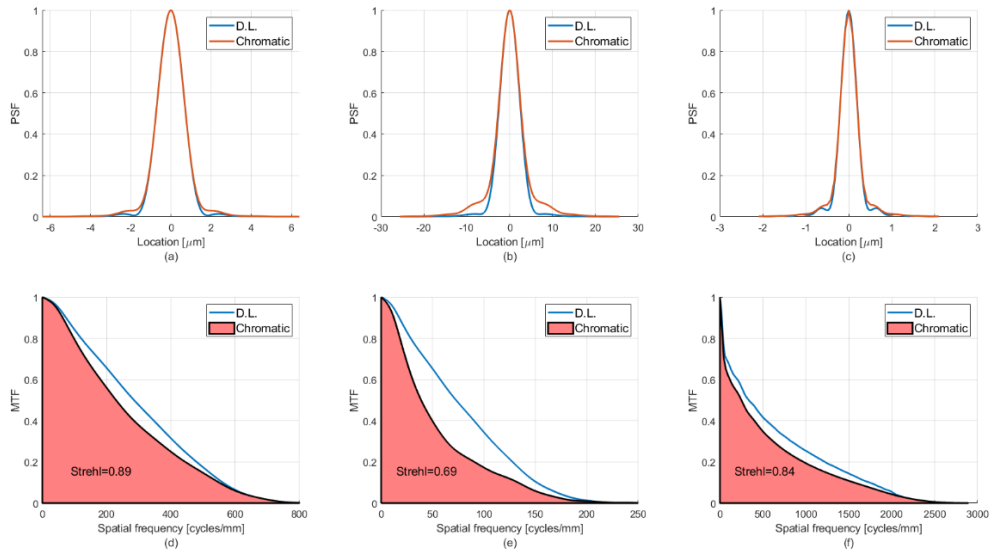

Fig. 4. PSFs, MTFs and 1D Strehl ratio of CDL designs equivalent to broadband AML and ADL designs of [37], compared to the diffraction limit. (a) PSF cross section of CDL equivalent to Design 1 of [37] (b) The same for Design 2 (c) The same for Design 3 (d) MTF of CDL equivalent to Design 1 of [37] (e) The same for Design 2 (f) The same for Design 3.

Table 2. Comparison of ADLs of [37] and AMLs quoted therein to equivalent CDLs

| Design | 1 | | | 2 | | | 3 | | |
|---|---|---|---|---|---|---|---|---|---|
| λmin [μm] | 0.47 | | | 3 | | | 0.56 | | |
| λmax [μm] | 0.67 | | | 5 | | | 0.8 | | |
| NA | 0.2 | | | 0.36 | | | 0.81 | | |
| EFL [μm] | 63 | | | 155 | | | 2 | | |
| Dia. [μm] | 25.7 | | | 119.6 | | | 5.5 | | |
| Fresnel no. | 4.6 | | | 5.6 | | | 4.1 | | |
| Airy rad [μm] | 1.74 | | | 6.78 | | | 0.51 | | |
| Chr. rad [μm] | 2.3 | | | 15 | | | 0.5 | | |
| DL FWHM | 1.48 | | | 5.26 | | | 0.37 | | |
| | AML [8] | ADL | CDL | AML [38] | ADL | CDL | AML [39] | ADL | CDL |
| Efficiency | 0.5 | 0.81 | 0.67 | 0.7 | 0.86 | 0.54 | 0.69 | 0.7 | 0.08 |
| FWHM | 1.5 | 1.31 | 1.47 | 5 | 5.49 | 5.46 | 0.39 | 0.39 | 0.36 |
| 2D Strehl | 0.92 | ? | 0.82 | 0.38 | ? | 0.5 | ? | ? | 0.62 |
| 1D Strehl | ~0.96 | 0.81 | 0.89 | ~0.69 | 0.55 | 0.69 | 0.29 | 0.29 | 0.84 |
| OPM | 0.68 | | 0.73 | 0.32 | | 0.51 | | | 0.24 |
| EOPM | 1.83 | | 1.96 | 1.25 | | 2.00 | | | 0.59 |

It should be noted that for the AML of Design 3 no claim is made that the design is achromatic [39], i.e. it is actually a conventional metalens that is simply used over a broad spectral band, with good results. It can seen from Fig. 4(f) why this is, since a CDL with identical parameters also gives near diffraction limited results over this range. It should also be noted that for the small aperture of design no. 3 (5.5 micron) our simulated CDL performance based on the scalar approximation may not be accurate. Such a design can be accurately simulated using FDTD or RCWA software. At any rate, lenses of such small dimensions are generally not relevant to imaging applications, which is the focus of this work.

The previous designs were analyzed assuming they were not intended for color imaging. For the case of color imaging the contribution of achromatic flat lenses becomes greater. This is because if the lens is optimally focused for the green (G) channel, the red (R) and blue (B) channels will be quite defocused. In the G channel, the central in-focus wavelengths give rise to a sharp peak in the PSF, which in turn creates a long "tail" in the MTF, allowing one to resolve high-frequencies (albeit with reduced contrast). As opposed to the G channel, the R and B channels may not have such a peak, since the image is defocused at these wavelength ranges. However, if there is an overlap in the spectral responsivities of the RGB channels (as typically occurs in a Bayer filter, see also Fig. S2), there may still be some sensitivity of the R and B channels at 550nm (center of the G channel), giving rise to a sharp PSF peak there too. Of course, the peak will be lower, because of the reduced sensitivity. In the section 8 of the Supplementary we compare results for several published color imaging AFLs to simulated CDL results.

## 6. Conclusion

The main goal of this paper is to facilitate future progress in the field of AFLs by introducing appropriate performance metrics and stressing the importance of comparing AFLs to the benchmark equivalent CFL. A secondary objective is to assess the potential of the various types of achromatic flat lenses based on these metrics, as an aid to researchers in their attempt to channel their research in the best direction.

Remarkably, the single channel (not color) broadband AMLs we have analyzed in this paper have worse overall imaging performance than the equivalent CDL. As for the ADLs we analyzed, it is difficult to determine their true performance, because a proper evaluation of their resolution was not made (only FWHM criterion was used). However, we suspect that given a proper evaluation, they would still not show an improvement compared to their CDL counterparts [16]. We note that in all analyzed cases, the overall transmission and veiling glare were not measured, and we assumed that the only transmitted light is that of the design order, so the calculated AFL OPMs are optimistic.

For color imaging, examples of which were analyzed in section 8 of the Supplementary, AFLs seem to have greater potential for improvement over CFLs. This is because in color imaging, the resolution of the R and B channels is greatly reduced for a CFL, precluding attainment of high color fidelity, while with AFLs the R and B channels can have resolution like the G channel, allowing high color fidelity. Still, for the AFLs analyzed in section 8 of the Supplementary, the overall performance did not exceed that of a CFL.

What is currently limiting the performance of spatially multiplexed AML designs (an example of which was analyzed in section 8.3 of the Supplementary) is low efficiency because of material absorption in silicon. The ideal solution would be a high-index material that is transparent in the visible range, which can allow high efficiency and low coupling between channels to be obtained. The problem is that there is no such material with refractive index as high as silicon. Promising candidates such as $TiO_2$ and GaN have not yet been applied to spatially multiplexed AMLs (GaN has been applied to a color-routing metalens, but there seems to be an issue of crosstalk between channels [40]).

Another important issue with flat lens performance, which we did not explore in this paper (although we did propose a metric for it in section 4), is the ability to retain the performance over a significant FOV. This has been addressed by using a removed aperture stop [41–43] and more recently using a phase induced effective aperture [44]. Regarding the latter method, we believe that an analysis based on the method presented in this work will show its limited usefulness, since despite the small FWHM, the Strehl ratio is very low, because of large spherical aberration.

We look forward to seeing new applications and improved performance in the field of flat lenses. We believe that progress in this field, enabling a transfer of achromatic flat lens technology to industry, necessitates not only innovative physics, but also down-to-earth good engineering practice. There are many advanced designs that can be achieved, and several additional degrees of freedom that can be utilized. We hope to see such implementations soon.

## Funding

Israel Ministry of Science, Technology and Space.

## Disclosures

The authors declare no conflicts of interest.

# Supplementary information for:

# A True Assessment of Flat Lenses for Broadband Imaging Applications

## 1   Diffraction efficiency vs. focusing efficiency

In this section we explain how the diffraction efficiency $\eta$ should be simulated and measured to obtain correct results, which can be used in our merit function. The concept of diffraction efficiency is well known and for the case of a lens is defined as the fraction of the incident energy (within the lens aperture) that goes to the design diffraction order [1–3], usually the first order. However, in recent years a new performance metric known as "focusing efficiency" has come into use in the flat lens community. To the best of our knowledge, the first to introduce this metric was Arbabi et al. in [4]. The focusing efficiency is defined there as the fraction of the incident energy that reaches the image plane region within a radius of 3×FWHM, centered on the center-of-mass of the focal spot.

To our understanding the intention of the authors of [4] was to define a measurable quantity that will closely represent the diffraction efficiency. Apparently, they assumed their lens produces near-diffraction-limited resolution, i.e., the focal plane intensity distribution of the first diffraction order is close to an Airy pattern. For an Airy pattern spot, 93.7% of the energy is contained within a radius of 3×FWHM, so this is a reasonable approximation of the diffraction efficiency. In other publications different radii were used in the definition of focusing efficiency, as indicated in [5].

The radius used for the focusing efficiency calculation is usually given in terms of the FWHM (e.g., 3×FWHM), and not of the Airy radius, despite their being quite similar for a diffraction limited lens. We conjecture the reason for this is that since the FWHM is related to the actual size of the central disk of the focal spot, it was thought that an area with a radius several times larger than the diameter of this disk will likely encompass almost all the energy, even in the case of a lens whose resolution is far from the diffraction limit. However, this is not usually true. As long as the aberrations are small enough that the Fraunhofer approximation holds [6], the FWHM represents the diameter of the central disk, and is fairly constant. However, these aberrations cause more energy to be directed to distant sidelobes, which may still fall outside the measured radius. If the aberration is larger than the Fraunhofer limit, there may be no central disk, so the FWHM is quite arbitrary, and 3×FWHM certainly cannot be assumed to contain almost all the energy.

To summarize, focusing efficiency, when defined as the fraction of incident energy contained within a radius of several FWHM at the image plane, can give a reasonable estimate of the diffraction efficiency if the lens has near diffraction limited resolution. However, if the lens is not nearly diffraction limited, it will give a result which represents a mixture of two effects: (a) Wavefront distortion of the design order, causing resolution degradation. (b) Diffraction efficiency, causing loss of signal and veiling glare, which results in SNR degradation. The weight of the contribution of each of the effects will depend on the radius of the chosen image plane region. Unless this type of measurement exactly matches the application (such as coupling to a multimode fiber with a core radius equal to the radius of the measurement region) such a measure of performance is not useful. Therefore, if one wants to measure true efficiency, a different method should be found for defining the region in the image plane over which the energy is collected. How should this region be defined?

The answer to this question is quite simple. The signal should be collected from the region over which the PSF is still varying, until it has "flattened out". For example, for the CDL PSF of Fig. 3(a) of the main text, this happens at about 10µm from the center. In Fig. 3(b) it happens at about 20µm, and in Fig. 3(c) it does not happen within the range shown in the graph. This is the **same region** that must be used for simulation/calculation of the MTF based on the Fourier transform of the PSF (see following section of this Supplementary for definition of MTF and PSF. As a note, the MTF of Fig. 3(f) was calculated based on a larger region than that shown in the PSF graph of Fig. 3(c)). Since almost all reports of flat lenses present PSF/MTF results, the size of this region must be known. All that now remains is to measure the efficiency over this same region.

Note that this measurement/simulation region does not necessarily correspond to 3×FWHM, 10×FWHM, etc., since it depends on the shape of the PSF. Of course, when evaluating the efficiency, it is also imperative to compare the "signal" to the **overall incident light**, and not only to a small part of it that happens to fit into the camera's active area or into the simulation window (as was erroneously done in [5]). A convenient way of performing an efficiency measurement using the same setup used to measure PSF/MTF, without the need for a pinhole, is described in [7].

## 2  Resolution metric – Strehl ratio vs. FWHM

It has long been recognized by the ultra-short laser pulse community that although the FWHM historically became popular as a measure of pulse temporal width, is not a good choice for most cases, since the pulse may have temporal substructure or broad wings causing a considerable part of the energy to lie outside the 50% intensity range. The metrics of choice are, therefore, the standard deviation of the pulse (when regarded as a probability density function) or the width of some type of autocorrelation function [8]. The equivalent problem occurs in the realm of spatial resolution, where FWHM seems to be a popular metric, although it is a very poor measure of resolution.

Examples of the problem with FWHM in the realm of spatial resolution are demonstrated in Fig. S1, for aberrated focal spots resulting from chromatic and spherical aberration of a conventional diffractive lens (CDL). The FWHM is compared to the widely accepted resolution metric of modulation-transfer-function (MTF), which is a function of spatial frequency [9], and to the Strehl ratio, which is a single number measure of how close the performance is to the diffraction limit. As can be seen in the figure, the FWHM fails to give any indication of the tremendous drop in resolution as a result of these aberrations (in fact, in our example the FWHM improves as a result of the spherical aberration, by an effect similar to that found in super-oscillation lenses [10]).

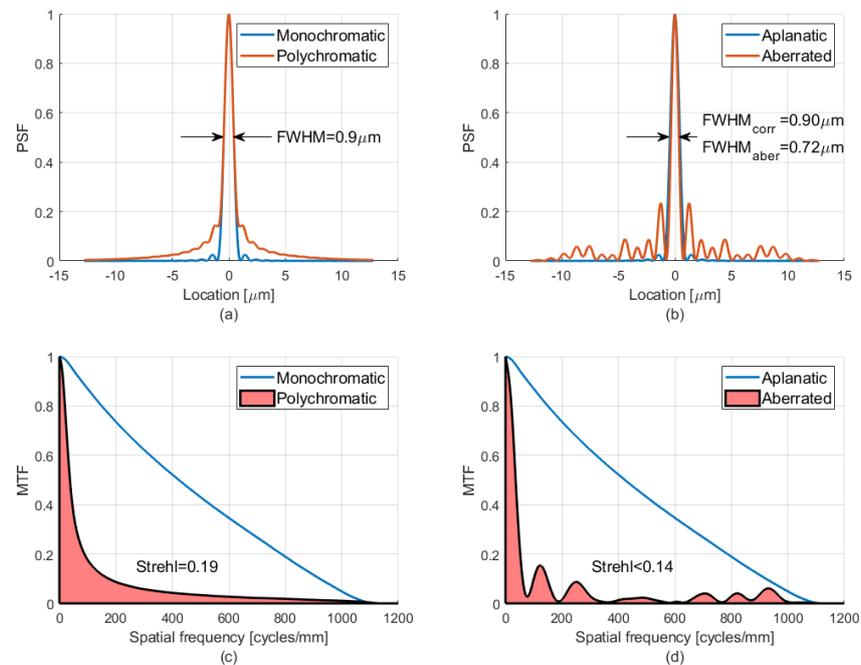

*Figure S1: FWHM resolution metric compared to MTF and Strehl ratio. The comparison is performed on a CDL of 3.36mm focal length, F/1 (NA 0.45), operating at wavelengths around 800nm. (a) PSF cross sections for monochromatic (diffraction limited) case and 20nm bandwidth (top-hat shaped spectrum) case. (b) PSF cross sections for monochromatic diffraction limited operation and the for the same lens with the two highest phase coefficients removed – thus introducing spherical aberration. (c) MTFs for PSFs of 'a'. (d) MTFs for PSFs of 'b'. The Strehl coefficients shown are the 1D version, equal to the pink shaded area under the MTF graph for the chromatic aberration case, and smaller than it for the spherical aberration case, since the OTF is negative for some frequencies.*

MTF, on the other hand, is the industry-standard spatial resolution metric for imaging applications. The MTF is the modulus of the optical-transfer-function (OTF), which in turn is the Fourier transform of the point-spread-function (PSF), the spatial intensity profile of the focal spot. The MTF gives the signal attenuation at each spatial frequency. It is analogous to the spectral amplitude of a laser pulse in the temporal realm.

Interestingly, the spectral intensity or amplitude is considered a poor indicator of temporal pulse width, since it lacks the phase information, which can cause significant pulse broadening. However, in the spatial realm, the MTF is considered a good spatial resolution metric, even though the phase information (given by the phase transfer function – PTF) is lost by the modulus operation. It seems that the reason for this is that while in temporal pulses it is quite common to have dispersion that affects the phase but not the amplitude of the spectral components, common lens aberrations will affect the amplitude as well as the phase of the OTF. Therefore, while the phase component of the OTF can have significant effect on image quality [9], a situation where the MTF is high, but the PSF is broad because of the PTF, is not realistic. Note that as opposed to the temporal realm, where the phase is difficult to measure, it is quite easy to obtain the PTF since it is simply the angular component of the complex OTF. Still, this information is usually ignored.

The Strehl ratio is defined as the ratio of the PSF peak to the diffraction limited PSF peak (when the PSFs are normalized so that the area under the PSF graphs is equal to 1). Because of the Fourier relation between MTF and PSF, the Strehl ratio is equal to the volume under the 2D MTF graph, relative to the volume under the diffraction limited 2D MTF (for real and positive OTF). The Strehl ratio is an optimal measure of performance for many non-imaging applications and certain imaging applications, such as astronomy or microscopy, where one needs to detect point-objects above the noise level. However, it is not ideal for typical wide-scene imaging applications. This is because the Strehl ratio is related to the volume under the 2D MTF, so high spatial frequencies are unduly given higher weight than the low spatial frequencies. Therefore, for most imaging applications it is more appropriate to use the one-dimensional (1D) version of the Strehl ratio, which is defined as the ratio of the line-spread function (LSF) peak to the diffraction limited LSF peak (the LSF is the one-dimensional version of the PSF obtained by summing the 2D PSF in one direction). As a result of the Fourier relation between the LSF and the 1D MTF, the LSF peak is equal to the area under the 1D MTF graph, for real and positive OTF [11,12]. All the examples shown in this paper are on-axis, therefore they have a symmetrical LSF, and therefore a real OTF. If there are no zeros in the OTF up to the frequency cutoff, the OTF is also positive. This is the case for all the examples in the paper, except that shown in Fig. S1(d), therefore we indicated there that the Strehl ratio is smaller than the area under the MTF and not equal to it.

Note that even under the condition of real and positive OTF, the relation between LSF peak and MTF is valid only if the peak of the LSF/PSF occurs on-axis. For an on-axis object, this will be the case if the aberrations are small enough for the Fraunhofer approximation to be accurate. If this is not the case, it is better to define the 1D Strehl ratio as the ratio of the areas under the MTFs, which is the convention we followed in this paper (this can be useful also for the off-axis case, where usually the PSF/LSF is not symmetrical, so again the peak may be off-axis). Usually, the LSF and MTF are calculated in horizontal and vertical directions, so we have horizontal and vertical 1D Strehl ratio. However, in this paper we deal only with radially symmetric systems, so there are single MTF and Strehl ratio values.

To summarize this section, we are not saying that the FWHM metric should never be used. It may be suitable for certain applications, such as evaluation of super-oscillation lenses for super-resolution imaging. However, this is the exception rather than the rule. For a conventional imaging lens, the FWHM is a very insensitive parameter, related primarily to the system central wavelength and relative aperture, and having little relation to the optical performance. The preferred single number performance parameter is the 1D or 2D Strehl ratio, depending on the application.

## 3   Zero frequency contrast measurement

As mentioned in section 2 of the main text, background light from spurious diffraction orders not only does not contribute to the signal but adds noise. Therefore, evaluating the level of background light is an important part of flat lens performance evaluation. The light from spurious diffraction orders is highly defocused, and therefore exists in the area beyond where the PSF flattens out. This light contributes to overall contrast reduction, i.e., it multiplies the entire MTF by a constant factor, which can be approximated by the $\eta/T$ factor mentioned in section 2 of the main paper [2]. This factor is equal to the contrast of very low spatial frequencies (since it is equal to the value of the un-normalized MTF at zero frequency). Note that we use the terms contrast and modulation interchangeably here, both defined by Eq. S2.

How can this factor be measured? It cannot be measured as part of the MTF, because the field-of-view (FOV) of the PSF measurement setup is too small (to obtain adequate spatial sampling of the PSF, large magnification is used, thus the FOV at the image plane is small). Therefore, the common engineering practice is to measure PSF/MTF by looking at the region until where the PSF flattens out, and to normalize the MTF so that at the zero frequency it is equal to 1. The zero-frequency contrast is relegated to a sperate measurement.

An indirect method of measuring the zero-frequency contrast, based on the $\eta/T$ approximation, is by placing a large detector immediately following the lens, to collect all the transmitted light. The ratio of this signal to that obtained from the incident light (using the same lens aperture) will give $T$, while $\eta$ is obtained from the previously discussed efficiency measurement (see section 1 of this Supplementary).

The direct method of measuring low spatial frequency contrast is known as veiling glare (VG) measurement. This measurement is performed with the lens coupled directly to a camera located at the image plane. A target composed of a large dark area on white background is viewed by the system. The veiling glare is then defined as [13]:

$$VG \equiv \frac{black\ level}{white\ level} \quad (S1)$$

Where the black and white level are defined relative to the 'capped black' level of the camera, which is the average signal level output by the camera when no light is incident.

The modulation of a signal is defined as:

$$M \equiv \frac{white\ level - black\ level}{white\ level + black\ level} \quad (S2)$$

By combining Eq. S1 and S2 we obtain Eq. S3, which gives us the low frequency modulation, as a function of the measured VG (one can of course just use Eq. S2 directly, and skip Eq. S1 and S3).

$$M \equiv \frac{1 - VG}{1 + VG} \quad (S3)$$

As explained, when the MTF is measured or calculated, it is generally normalized so that at zero frequency it is equal to 1. The parameter $M$ represents the absolute (un-normalized) MTF value at zero frequency, via Eq. S2 or S3. Therefore, we can multiply the MTF by this factor, to obtain the absolute MTF. Despite this possibility, it is customary to separate the two effects, i.e., use the normalized MTF and the VG as two separate metrics of system performance, the first representing the system 'resolution', and the second representing the system low frequency 'contrast'. Note that for our metric we need both $\eta$ and $T$, so it is not enough to measure only VG. We must measure at least 2 out of these three parameters, preferably $\eta$ and VG (from the VG we can obtain $M$, and from $M$ and $\eta$ we can obtain the "true" effective $T$ that contributes to the background signal).

The tricky part about VG measurement is that the level of background radiation depends on the ambient illumination. Therefore, one must attempt to match the measurement setup to the application, and of course perform functional tests as well. However, if the VG is stemming from spurious diffraction orders, and not from parasitic reflections from mechanical housing and optical surfaces, as is usually the case for flat lenses, it should be less sensitive to ambient illumination conditions.

## 4 Derivation of color EOPM

As explained in the main paper, to obtain good color fidelity we need to have good SNR and resolution, represented by the EOPM, for each one of the RGB channels. This means that in addition to requiring high average EOPM for the three channels, we need to require low standard deviation among them.

The average EOPM is given by:

$$avg(EOPM) = \frac{EOPM_R + EOPM_G + EOPM_B}{3} \quad (S4)$$

The standard deviation is given by:

$$std(EOPM) = \sqrt{\frac{(EOPM_R - avg)^2 + (EOPM_G - avg)^2 + (EOPM_B - avg)^2}{2}} \quad (S5)$$

For a case where we have non-zero EOPM in only one channel, say the green channel:

$$EOPM_R = EOPM_B = 0 \Rightarrow avg(EOPM) = \frac{EOPM_G}{3}$$
$$\Rightarrow EOPM_G = 3avg(EOPM) \quad (S6)$$

Substituting Eq. S6 into Eq. S5 we obtain:

$$std(EOPM) = \sqrt{3}\ avg(EOPM) \quad (S7)$$

Therefore, we choose the color merit function to be the average EOPM, multiplied by the correction factor $1 - std(EOPM)/\sqrt{3}\ avg(EOPM)$. This factor is equal to 1 when the standard deviation is zero, i.e., maximum color fidelity, and 0 when the standard deviation is as calculated above for the case of signal from only one channel.

## 5   Derivation of extended FOV EOPM

The information content of an image, in bits per unit area, is given by [14]:

$$C = \iint_{-\infty}^{\infty} \log\left[1 + \frac{MTF^2(\nu_x, \nu_y) \cdot P(\nu_x, \nu_y)}{N(\nu_x, \nu_y)}\right] d\nu_x d\nu_y \qquad (S8)$$
$$= \iint_{-\infty}^{\infty} \log[1 + MTF^2(\nu_x, \nu_y) \cdot SNR^2(\nu_x, \nu_y)] d\nu_x d\nu_y$$

Where $P$ is the signal power spectral density, and $N$ is the noise power spectral density. Note the similarity between Eq. S8 and Eq. 1, in that we have the product of the MTF and SNR. The main difference is *log* function that is applied before the integration is carried out. It is tempting to neglect the 1 inside the *log*, since usually the SNR will be much larger than 1. However, this cannot be done, since at higher spatial frequencies the MTF will go down towards zero. The 1 inside the *log* is necessary for these high spatial frequencies, since otherwise the *log* will take on negative values, that will incorrectly reduce the amount of information, instead of only not adding to it. We assume "white" signal and noise within the system bandwidth, so the SNR is constant, giving us Eq. S9:

$$C = \iint_{-\nu_{co}}^{\nu_{co}} \log[1 + SNR^2 \cdot MTF^2(\nu_x, \nu_y)] d\nu_x d\nu_y \qquad (S9)$$

Where $\nu_{co} = 2NA/\lambda$ is the diffraction limit cutoff frequency of the lens. Assuming a shot noise limited system, the SNR can be calculated according to [15]:

$$SNR = \left[\frac{A_{pix} t}{hc} \int \pi QE(\lambda) \cdot \lambda \cdot L_\lambda(\lambda) \times \eta(\lambda) \times (NA)^2 d\lambda\right]^{1/2} \qquad (S10)$$

Where $A_{pix}$ is the pixel area, $t$ is the camera integration time, $h$ is Planck's constant, $c$ the speed of light in vacuum, $QE$ the camera quantum efficiency, $L_\lambda$ the object spectral radiance, $\eta$ the flat lens efficiency, and $NA$ is the numerical aperture of the lens. To simplify, we assume the spectral range is small enough so that the various parameters in the integral can be considered constant. We then obtain:

$$SNR = NA\left[\frac{A_{pix} t}{hc} \pi QE \cdot L_\lambda \cdot \eta \cdot \lambda \cdot \Delta\lambda\right]^{1/2} \qquad (S11)$$

The above derivation assumes no background illumination that contributes to noise but not to signal. To include the background noise, we must replace $\eta^{1/2}$ in Eq. S11 with $\eta/\sqrt{T}$. When using this merit function, we must compare based on absolute differences, not relative, since the *log* function converts products into sums. To obtain the overall information we multiply the information per unit area, $C$, by the image area $A$:

$$EOPMfov = A \cdot C \qquad (S12)$$

The drawback of this merit function is that it is not possible to separate the lens from the camera and the scene illumination. So, to compare between types of flat lenses, we must consider the specific application. Regarding the image area, we can compare lenses over equal image areas, without loss of generality. Since typically the MTF will change over different image regions, we suggest breaking up the image into several areas (rings in the case of radially symmetric systems) and summing the merit functions for the different areas. If the camera has not yet been determined, one can make reasonable assumptions about it for use in the calculation (such as $QE=1$, $t$ based on the desired frame rate, reasonable pixel size based on the state of the art, etc.). However, one must have an estimate of the ambient lighting since it too will affect the choice of lens aperture.

# 6 Commercial optical design software simulation

In the simulations shown in this paper, the MTF and Strehl ratios are polychromatic, i.e., calculated over a certain spectral range. The optical design software performs polychromatic simulations by simulating many discrete wavelengths, defined by the user, and then performing a weighted average, according to weights defined by the user. We made sure to use enough wavelengths to simulate a continuous spectrum.

All the MTF simulations shown in the paper are for a diffractive surface. The radially symmetric 'Binary2' surface type was used in Zemax for all the simulations. For the extended depth of focus (EDOF) metalens (section 8.4 of this Supplementary), a non-radially symmetric 'Binary1' surface was added, to describe the cubic phase.

All the PSFs and MTFs in this paper were calculated using the Huygens PSF and MTF option in Zemax. This was necessary since the aberrations are large in many of the cases, to the level where the Fraunhofer approximation is no longer valid in the image plane [6].

For high-NA lenses the computation of the PSF/MTF is more complicated since the Z-component of the electric-field must be considered [16–18]. In addition, in the case of a flat lens we obtain higher angular ray density, and therefore higher intensity of light, at the edges of the aperture. This creates a pupil apodization effect, like that of a central obscuration [19]. While Zemax can account for these effects, we opted to use Code V for the high-NA cases, since we found a difference between the two softwares, and our impression is that Zemax is overemphasizing the pupil apodization (this subject is still under investigation). Therefore, the diffraction limited MTFs of Fig. 4(f) and 5(f) do not have the typical shape of low NA diffraction limited MTFs. We used input circular polarization in Code V to simulate unpolarized light.

# 7 CDL efficiency calculation

An analytic approximation of the first order diffraction efficiency for an *N* level diffraction grating, at the design wavelength and normal incidence, is given by [20]:

$$\eta = \frac{4n}{(n+1)^2} sinc^2(1/N) \left(1 - 2\frac{N-1}{N}\frac{1}{n-1}\frac{\lambda}{d}tan(\theta)\right) \tag{S13}$$

Where $n$ is the substrate refractive index (which we took as 1.5 for our calculations), $\lambda$ is the wavelength of light, $d$ is the grating period, $\theta$ is the diffraction angle (assuming normal incidence), and $sinc(x) = \frac{sin(\pi x)}{\pi x}$. The first factor (before the *sinc*) accounts for Fresnel reflection from the diffractive surface, the *sinc* accounts for the effect of the phase sampling, and the last factor (in parentheses) accounts for the shadowing effect. This approximation is limited to $\lambda/d$ that is not too small - for first order of diffraction it should be larger than about 2. For smaller values of $\lambda/d$ (which cause the approximate result to be negative of complex), I took the pessimistic estimate of zero efficiency.

For the case of a diffractive lens, the local period changes along the radial direction of the aperture. Assuming a certain minimum feature size $\Delta$, we can calculate the local number of phase levels, $N$, and therefore the local shadowing (the local period and diffraction angle are calculated using the grating equation). Integration over the aperture is then carried out to obtain the average efficiency over the full aperture. For the cases where the CDL is compared to an ADL, we used the same feature size for the CDL as was used in the published ADL (1µm, 4µm, and 0.35µm for designs 1, 2, and 3 of Table 2 respectively, and 3µm and 1.2µm for designs 1 and 2 of Table S2 respectively). For the cases where the CDL is compared to an AML, we used a feature size of 1µm.

In addition to the above efficiency calculation, we must also account for the efficiency degradation because of wavelength detuning. The first order diffraction efficiency averaged over a spectral range of $\Delta\lambda$, with uniform weighting, centered on the design wavelength $\lambda_0$ is given by [21]:

$$avg(\eta) \approx 1 - \left(\frac{\pi\Delta\lambda}{6\lambda_0}\right)^2 \tag{S14}$$

The product of these two expressions was used to estimate the efficiency of the CDL designs presented in the paper in Tables 1 and 2, with $\Delta\lambda$, $\lambda_0$ per design. The maximum number of phase levles is taken as 8, and the number is reduced to equal the integer part of $(d/\Delta)$ if it is less than 8.

For the color imaging examples presented in section 8 of this supplementary, the wavelength detuing efficiency was cacluated using the formula for diffraction efficiency as a function of wavelength, since Eq. S14 is limited to uniform spectral weight (for the first order of diffraction, neglecting the material dispersion) [22]:

$$\eta = sinc^2(\lambda_0/\lambda - 1) \tag{S15}$$

Using this formula, a weighted average, based on the spectral responsivity of the system, was performed. Note that strictly speaking, it is not correct to multiply the efficiency formulas based on number of levels/shadowing (Eq. S13) and based on wavelength (Eq. S14 and S15) to obtain the overall efficiency, since each assumes the other does not exist (level formula assumes single wavelength, and spectral formula assumes perfect blaze), when in fact there is coupling between the two effects. However, we neglected this effect.

Another point to mention is that low efficiency near the aperture edges will also affect the MTF. This would cause the diffraction limited MTF to drop, and the aberration blur to decrease, so the simulated MTF would be even closer to the diffraction limit (higher Strehl ratio), leaving less room for improvement because of chromatic correction. In our analysis we took the stringent assumption that the MTFs are not affected by the efficiency.

# 8 Achromatic flat lenses for color imaging
## 8.1 Dispersion engineered metalens

Here we analyze the performance of the AML presented in [23], based on dispersion engineered nanostructures, and compare to the performance of an equivalent CDL. This metalens has first-order parameters similar to Design 1 shown in Table 2 (originally published in [24]), but is improved in the sense that it is not polarization sensitive, and the spectral range is slightly extended. While in Table 2 we looked at performance over the entire spectral range, here we break up the spectral range into R, G and B channels, and look at the color imaging performance.

In Fig. S2 we show the spectral responsivity of the Thorlabs DCC1645C color camera, based on manufacturer data[1]. We used this RGB spectral responsivity in our analysis but assumed an added long-pass filter with cut-on at 450nm was added in front of the lens (grayed out area in Fig. S2), since some of the designs analyzed were not optimized for shorter wavelengths. In Fig. S3 the equivalent CDL performance, over the RGB spectral bands of the camera, at a single common focal plane, is shown.

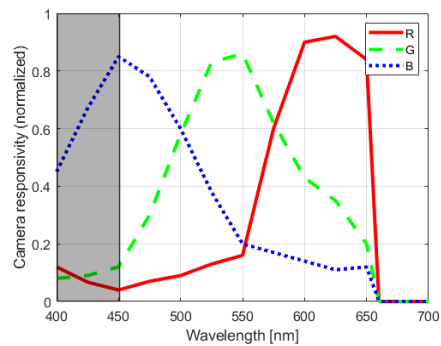

*Figure S2: Spectral sensitivity of RGB channels of Thorlabs DCC1645C color camera. There is significant spectral overlap between the channels.*

---

[1] https://www.thorlabs.com/newgrouppage9.cfm?objectgroup_id=4024&pn=DCC1645C#5316

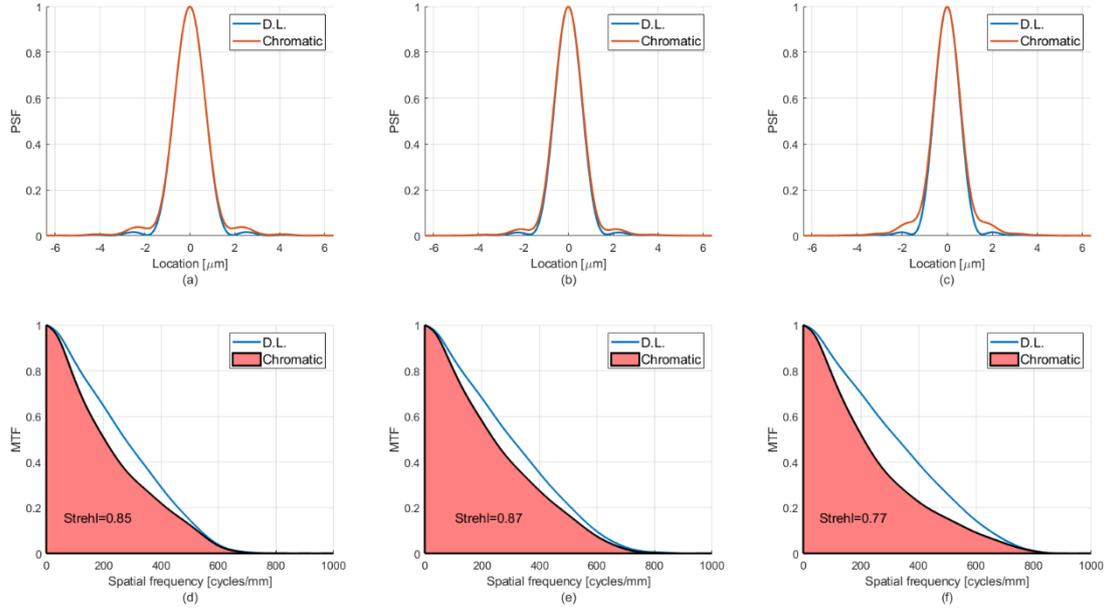

*Figure S3: PSFs and MTFs of chromatic diffractive lens, with EFL 67µm and NA 0.2, over RGB spectral ranges defined according to spectral sensitivity of Thorlabs DCC1645C color camera. (a) R PSF cross section (b) G PSF cross section (c) B PSF cross section (d) R MTF (e) G MTF (f) B MTF*

Table S1. Comparison of AML of [23] to equivalent CDL

| Design | 1 | | | | | |
|---|---|---|---|---|---|---|
| λmin [µm] | 0.45 | | | | | |
| λmax [µm] | 0.65 | | | | | |
| NA | 0.2 | | | | | |
| EFL [µm] | 67 | | | | | |
| Dia. [µm] | 27 | | | | | |
| Fresnel no. | 5 | | | | | |
| Airy rad [µm] | 1.7 | | | | | |
| Chr. rad [µm] | 2.5 | | | | | |
| | AML [23] | | | CDL | | |
| | R | G | B | R | G | B |
| Efficiency | 0.3 | 0.3 | 0.3 | 0.61 | 0.62 | 0.59 |
| 2D Strehl | 0.8 | 0.8 | 0.8 | 0.65 | 0.83 | 0.67 |
| 1D Strehl | 0.9 | 0.9 | 0.9 | 0.85 | 0.87 | 0.77 |
| OPM | 0.49 | 0.49 | 0.49 | 0.66 | 0.68 | 0.59 |
| EOPM | 1.50 | 1.37 | 1.24 | 1.67 | 1.90 | 1.80 |
| EOPMcolor | 1.29 | | | 1.72 | | |

It can be seen from Table S1 that despite the improved resolution provided by the AML, the overall performance (OPM, EOPM and EOPMcolor) of the CDL is better. This is because, for the first order design parameters of the AML, the CDL does not have severe chromatic aberration (because of the low Fresnel number). Therefore, the small reduction in resolution is more than compensated by the improved efficiency of the CDL.

## 8.2 Achromatic diffractive lens (ADL)

Here we analyze the performance of ADLs presented in [25], and compare to the performance of equivalent CDLs. We will look at two designs, one with NA of 0.05, and the other with NA of 0.18. In Fig. S4 we show the equivalent CDL performance for the first design (NA 0.05), over the RGB spectral bands of the Thorlabs DCC1645C color camera. In Fig. S5 we show the same for the second (NA 0.18) design. The performance comparison is summarized in Table S2.

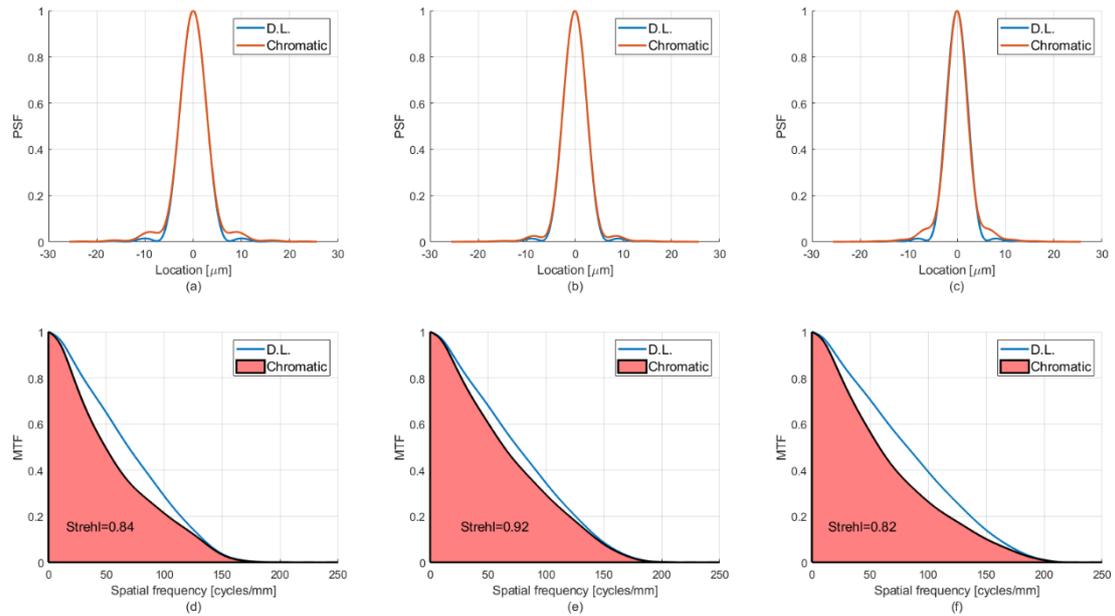

*Figure S4: PSFs and MTFs of CDL equivalent to design 1 (NA 0.05) of [25] over RGB spectral ranges, defined according to spectral sensitivity of Thorlabs DCC1645C color camera. (a) R PSF cross section (b) G PSF cross section (c) B PSF cross section (d) R MTF (e) G MTF (f) B MTF*

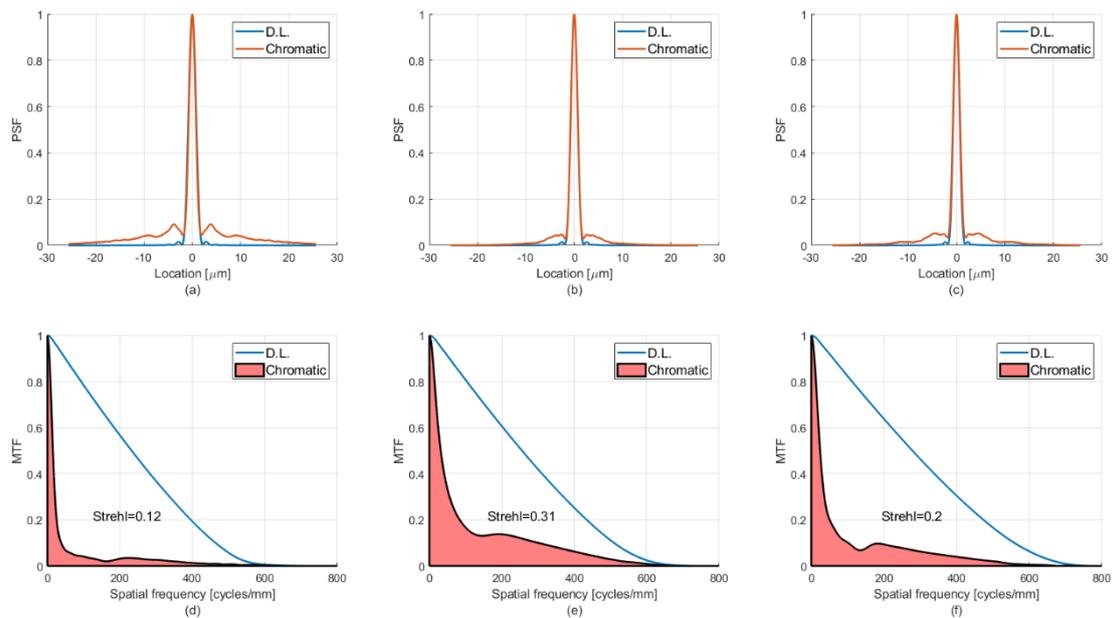

*Figure S5: PSFs and MTFs of CDL equivalent to design 2 (NA 0.18) of [25] over RGB spectral ranges, defined according to spectral sensitivity of Thorlabs DCC1645C color camera. (a) R PSF cross section (b) G PSF cross section (c) B PSF cross section (d) R MTF (e) G MTF (f) B MTF*

Table S2. Comparison of ADL of [25] to equivalent CDL

| Design | 1 | | | | | | 2 | | | | | |
|---|---|---|---|---|---|---|---|---|---|---|---|---|
| λmin [μm] | 0.45 | | | | | | 0.45 | | | | | |
| λmax [μm] | 0.65 | | | | | | 0.65 | | | | | |
| NA | 0.05 | | | | | | 0.18 | | | | | |
| EFL [μm] | 1000 | | | | | | 1000 | | | | | |
| Dia. [μm] | 100 | | | | | | 366 | | | | | |
| Fresnel no. | 4.6 | | | | | | 60.4 | | | | | |
| Airy rad [μm] | 6.71 | | | | | | 1.86 | | | | | |
| Chr. rad [μm] | 9.1 | | | | | | 33.3 | | | | | |
| DL FWHM | 5.5 | | | | | | 1.5 | | | | | |
| | ADL | | | CDL | | | ADL | | | CDL | | |
| | R | G | B | R | G | B | R | G | B | R | G | B |
| Efficiency | 0.3 | 0.5 | 0.4 | 0.58 | 0.59 | 0.56 | 0.2 | 0.2 | 0.2 | 0.58 | 0.59 | 0.56 |
| FWHM [μm] | 7 | 6 | 4 | 6.17 | 5.64 | 4.94 | 7 | 7 | 2 | 1.62 | 1.58 | 1.57 |
| 1D Strehl | 1.17 | 1.00 | 0.95 | 0.84 | 0.92 | 0.82 | 0.20 | 0.53 | 0.33 | 0.12 | 0.31 | 0.20 |
| OPM | | | | 0.64 | 0.70 | 0.61 | | | | 0.09 | 0.24 | 0.15 |
| EOPM | | | | 1.45 | 1.78 | 1.7 | | | | 2.75 | 7.93 | 5.49 |
| EOPMcolor | | | | 1.54 | | | | | | 3.89 | | |

Since the only resolution data given in [25] is the FWHM, we cannot properly asses the performance of the ADLs. The 1D Strehl ratio values appearing in the table in blue indicate what Strehl ratio the ADL would need to exceed the performance of the CDL. For design 1 the Strehl ratios are near or above 1, meaning that with the reported efficiencies there is no way the ADL can achieve better performance than the equivalent CDL. This is not surprising, since the MTFs of the CDL shown in Fig. S4 are close to the diffraction limit. For design 2, the CDL MTF performance shown in Fig. S5 is quite poor. Therefore, there is much room for the ADL to give improved performance. However, since only FWHM was reported, we cannot know if it does. Again, the 1D Strehl ratios needed for the ADL to exceed the CDL performance are shown in blue. This time they are well below 1, so they can potentially be achieved. However, based on the FWHMs of the ADL, which are larger than those of the CDL, it is not likely that they are achieved (it should be noted that FWHM significantly larger than diffraction limit is usually an indicator of pupil apodization caused by decreased efficiency near the aperture edge, and not of chromatic or spherical aberration, which as shown in Fig. 2 have little effect on the FWHM. The other option is that the aberrations are so large that the Fraunhofer approximation is not valid in the image plane, so we get a Fresnel diffraction pattern which can be broader).

### 8.3 Spatial multiplexed metalens

To demonstrate the performance of a spatially multiplexed achromatic metalens, we used the design presented in [26], based on a-Si nano-blocks, which was applied there to an achromatic hologram. Here we apply it to a metalens, with the same parameters as the second ADL design of section 8.2 of this supplementary (EFL 1 mm, NA 0.18). The nanoantenna spectral response, shown in Fig. S6, is narrower than the camera spectral response, shown in Fig. S2, thus dominating the response. We therefore used the nanoantenna spectral response for the wavelength weights in Zemax.

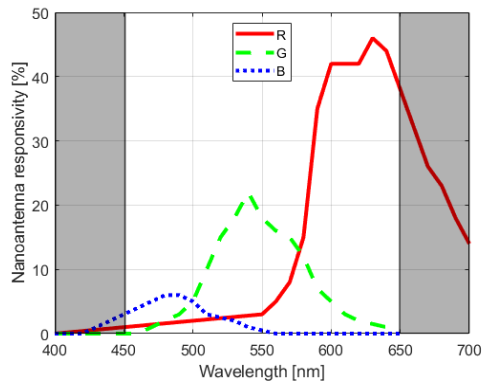

*Figure S6: Spectral response of nanoantennas of [26]*

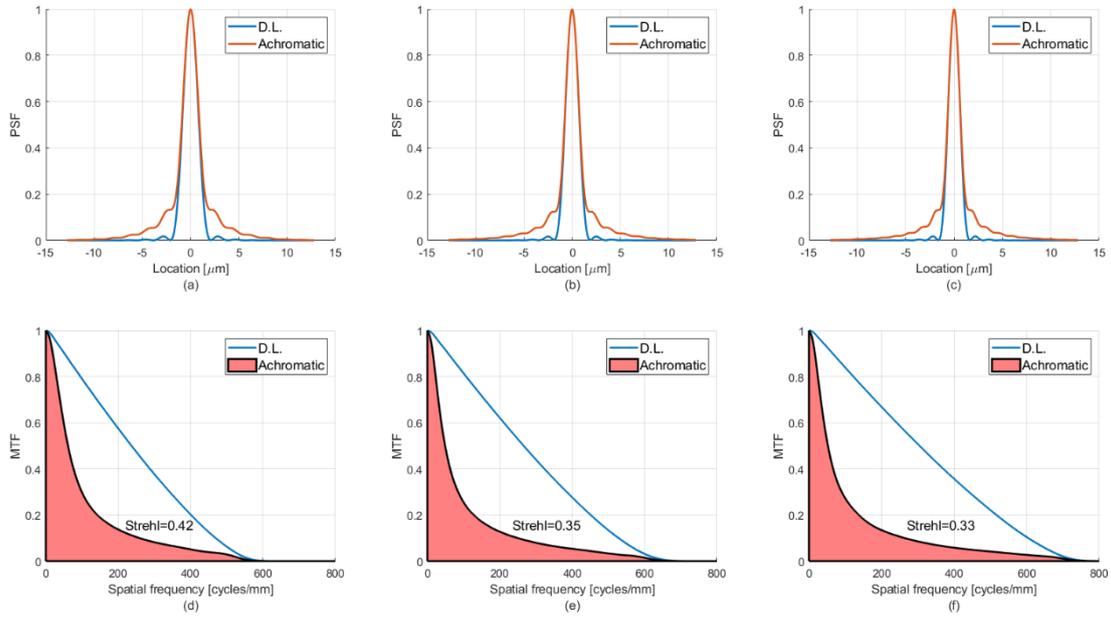

*Figure S7: PSFs and MTFs of spatial multiplexed AML over RGB spectral ranges, defined according to spectral response of nanoantennas given in [26], with lower and upper cutoffs at 450nm and 650nm respectively. (a) R PSF cross section (b) G PSF cross section (c) B PSF cross section (d) R MTF (e) G MTF (f) B MTF*

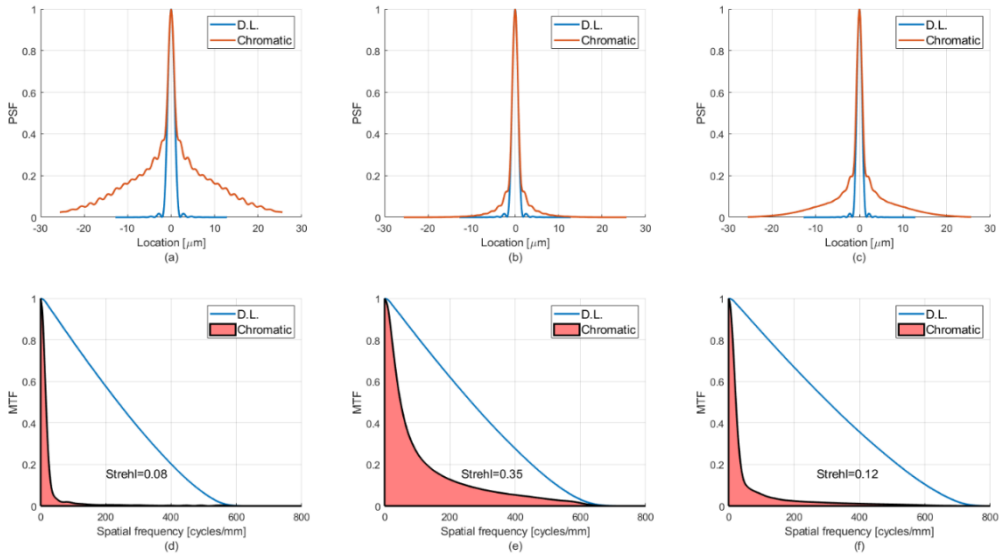

*Figure S8: PSFs and MTFs of CDL over RGB spectral ranges, defined according to spectral response given in [26], with lower and upper cutoffs at 450nm and 650nm respectively (a) R PSF cross section (b) G PSF cross section (c) B PSF cross section (d) R MTF (e) G MTF (f) B MTF*

In Fig. S7 the MTFs for the spatial multiplexed AML simulated in Zemax are shown. The simulation was performed by simulating a diffractive lens in Zemax and adjusting the phase function separately for each of the RGB channels, to obtain optimal focus at the common focal plane. In this manner we obtain similar performance for each of the RGB channels, with minor variations due to the differences in relative spectral width of the channels.

In Fig. S8 the MTFs for an equivalent CDL are shown. As expected, the G channel MTF is identical to that of Fig. S7. However, the B and R channels are now greatly defocused, giving rise to the low MTFs shown in Fig. S8(d) and (f). Despite this, the overall performance (*EOPMcolor*) of the CDL, shown in Table S3, is better than the AML. The reason for this is the low efficiency of the AML, especially at the shorter wavelengths (B and G channels), because of the absorption in the silicon nano-blocks.

Even though we do not see improved performance here, the spatial multiplexing method may have potential if the efficiencies can be improved by using a more transparent material. This was done in [27] using GaN nanoantennas. However, since the polarization conversion efficiencies are high over the entire spectrum for all three antenna types (R, G and B), it seems there is high crosstalk there. Another option is longitudinal cascading of metalenses, which was done in [28] using metallic nanoantennas. However, this design probably suffers from low efficiency. If done with dielectric nanoantennas, the challenge is to get each of the RGB layers to interact only with the relevant spectral range and be transparent to the other ranges.

Table S3. Comparison of AML based on nanoantennas of [26] to equivalent CDL

| Design | 1 | | | | | |
|---|---|---|---|---|---|---|
| λmin [µm] | 0.45 | | | | | |
| λmax [µm] | 0.65 | | | | | |
| NA | 0.18 | | | | | |
| EFL [µm] | 1000 | | | | | |
| Dia. [µm] | 336 | | | | | |
| Fresnel no. | 60.4 | | | | | |
| Airy rad [µm] | 1.86 | | | | | |
| Chr. rad [µm] | 33.3 | | | | | |
| | AML | | | CDL | | |
| | R | G | B | R | G | B |
| Efficiency | 0.15 | 0.05 | 0.02 | 0.58 | 0.59 | 0.56 |
| 1D Strehl | 0.42 | 0.35 | 0.33 | 0.08 | 0.35 | 0.12 |
| OPM | 0.16 | 0.08 | 0.04 | 0.06 | 0.27 | 0.09 |
| EOPM | 4.39 | 2.21 | 1.05 | 1.62 | 7.30 | 2.22 |
| EOPMcolor | 1.57 | | | 1.91 | | |

## 8.4 Extended depth of focus metalens

The idea behind this type of metalens is to add a cubic phase function to the basic focusing hyperbolic phase function, thus creating an extended depth-of-focus (EDOF). If the defocus resulting from chromatic aberration is smaller than the depth-of-focus, a focal plane where all wavelengths are in focus can be found. In [29] an EDOF metalens is presented for color imaging. Since the lens parameters are given in the paper, including the cubic phase coefficient, we were able to simulate the lens in Zemax. The monochromatic PSFs and MTFs obtained by our Zemax simulation for the three center-wavelengths used in [29] are shown in Fig. S9.

The PSFs are quite similar to the measured results shown in Fig. 2 E-G of [29], but the MTFs are quite different from the calculated results shown in Fig. 2H of [29]. We believe these calculated MTFs are in error, for several reasons: (1) If these MTFs are correct, the EDOF lens gives near diffraction limited resolution, so why is a deconvolution needed to obtain good resolution? (2) In [29] the blue and red images from the EDOF (Fig. 3D) do not look much better than the equivalent images from the conventional metalens (Fig. 3C). However, the red and blue calculated MTFs of the EDOF shown in Fig. 2H are much higher than those of the conventional lens, shown in Fig. 2D. (3) In Fig. 4A of [29], the resolution of the letters R and B looks better for the "Singlet" (conventional) metalens, than for the EDOF metalens. (4) The MTF cutoff stated in [29] in the legend of Fig. 2 is 579c/mm, when in fact it should be $v_{c.o.} = \frac{2NA}{\lambda} = \frac{2 \cdot 0.45}{530e-6} \approx 1700 c/mm$.

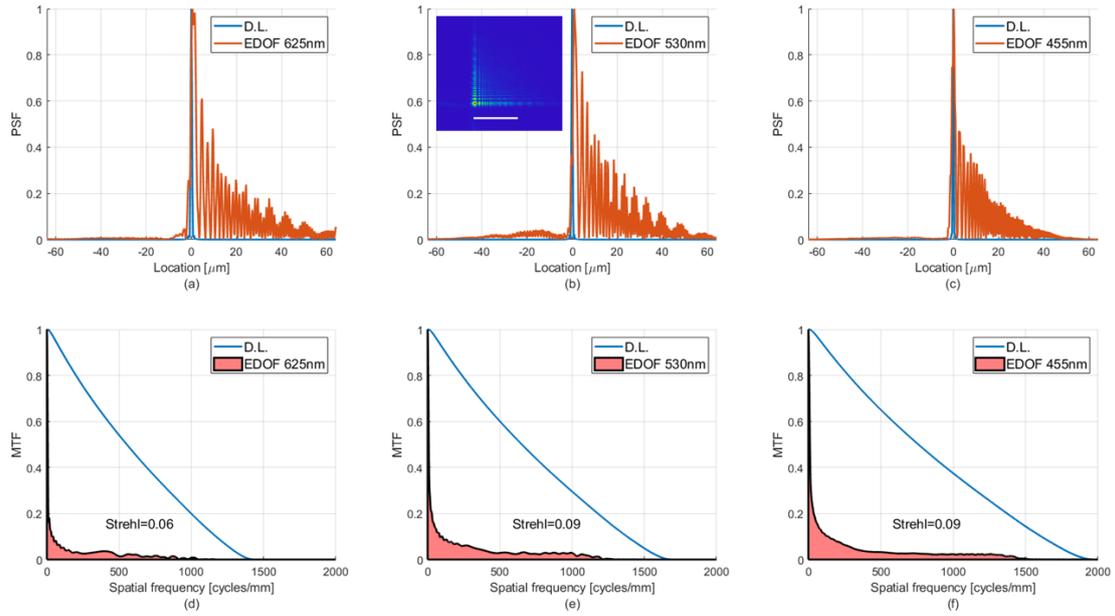

*Figure S9: PSFs and MTFs of EDOF lens at wavelengths used in* [29]. *(a) 625nm PSF cross section (b) 530nm PSF cross section. Inset - 2D PSF. Scale bar is 30µm (c) 455nm PSF cross section (d) 625nm MTF (e) 530nm MTF (f) 455nm MTF.*

In Fig. S10 we show the simulated MTFs for the same lens, but over the RGB spectral ranges of the Thorlabs color camera. The MTFs for the R and G channels (Fig. S10 (d-e)) are worse than those of the monochromatic wavelengths shown in Fig. S9 (d-e). This seems to be a result of the fact that, as can be seen in Fig. 1E of [29], the focal spot shifts in the transverse direction for different focal planes. This means that the same effect will occur at a single focal plane for different wavelengths. So, if we have polychromatic light, the overall MTF will be worse than the average MTF of the different wavelengths (The effect is accounted for by the overall OTF, which is the weighted average of the OTFs. The phase of the OTF accounts for the transverse shift of the PSFs). Therefore, this type of EDOF lens may not be a good method for creating an achromatic lens for imaging purposes.

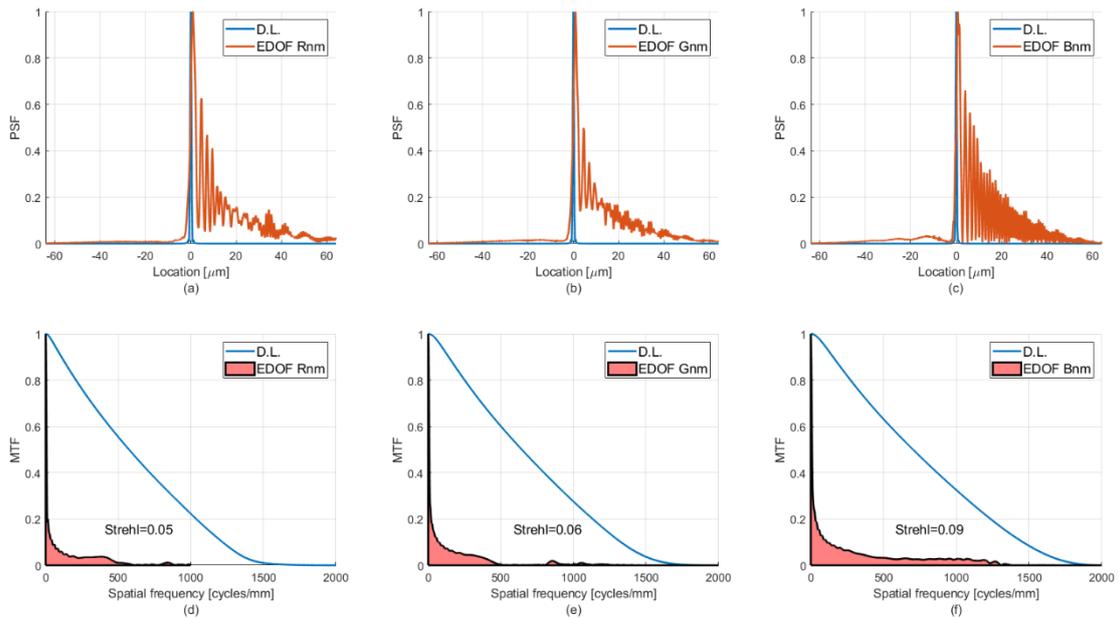

*Figure S10: PSFs and MTFs of EDOF lens over RGB spectral ranges, defined according to spectral sensitivity of Thorlabs DCC1645C color camera. (a) R PSF cross section (b) G PSF cross section (c) B PSF cross section (d) R MTF (e) G MTF (f) B MTF*

Last, but not least, in Fig. S11 we show the MTFs for an equivalent CDL. Poor as the MTFs are, because of chromatic aberration, they are still better than the EDOF MTFs of Fig. S10. As a result, the overall performance (*EOPM* and *EOPMcolor*) shown in Table S4 is better for the CDL. This is in line with the third point mentioned previously, that the R and B images shown in Fig 4A of [29] look better than those of the EDOF.

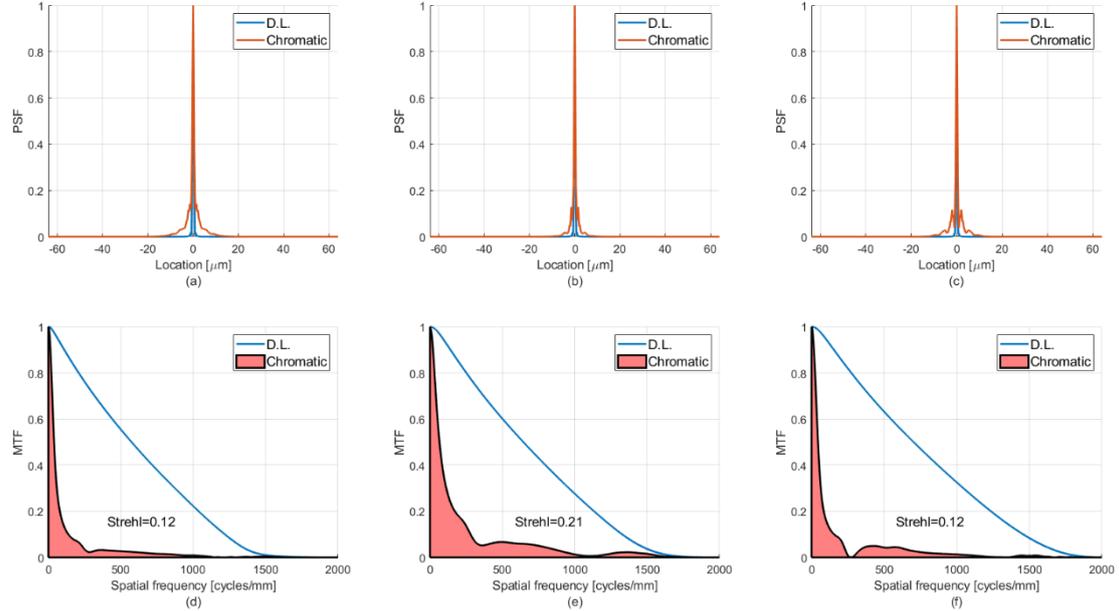

*Figure S11: PSFs and MTFs of chromatic diffractive lens (with parameters equivalent to EDOF lens) over RGB spectral ranges, defined according to spectral sensitivity of Thorlabs DCC1645C color camera. (a) R PSF (b) G PSF (c) B PSF (d) R MTF (e) G MTF (f) B MTF*

Table S4. Comparison of AML based on EDOF lens of [29] to equivalent CDL

| Design | 1 | | | | | |
|---|---|---|---|---|---|---|
| λmin [μm] | 0.45 | | | | | |
| λmax [μm] | 0.65 | | | | | |
| NA | 0.45 | | | | | |
| EFL [μm] | 200 | | | | | |
| Dia. [μm] | 200 | | | | | |
| Fresnel no. | 85.7 | | | | | |
| Airy rad [μm] | 0.75 | | | | | |
| Chr. rad [μm] | 18.2 | | | | | |
| | AML | | | CDL | | |
| | R | G | B | R | G | B |
| Efficiency | 0.85 | 0.85 | 0.85 | 0.58 | 0.59 | 0.56 |
| 1D Strehl | 0.05 | 0.06 | 0.09 | 0.12 | 0.21 | 0.12 |
| OPM | 0.05 | 0.06 | 0.08 | 0.09 | 0.16 | 0.09 |
| EOPM | 1.97 | 2.63 | 4.32 | 3.9 | 7.6 | 4.7 |
| EOPMcolor | | 2.27 | | | 4.27 | |

In a more recent publication [30] the authors implemented symmetric EDOFs, which solve the problem of the lateral shift of the PSF, and compared their results to conventional metalens results. The reported MTF results seem reasonable. However, in our opinion there is a flaw in the comparison, since it is assumed that the deconvolution kernel must be the same for the R, G and B images, when in fact there is no reason for this. Another difference between our simulation and that of [30] is that we assume a natural scene, so there is overlap in the spectral responses of R, G and B channels. They used an artificial image of an OLED monitor, so there is no spectral overlap. The lack of overlap degrades the conventional metalens resolution.

The SSIM metric used in [29,30] to evaluate overall performance is in our opinion a poor metric, since it lacks physical insight (for example, is the degradation due to noise or to blurring?) and requires a reference image. This metric is more suitable for evaluation of image compression algorithms. The statement made by the authors of [30] that the Strehl ratio is not relevant to their lens, since they perform a deconvolution, is incorrect. The reason deconvolution is useful is because the human brain is much better at averaging noise than it is at performing deconvolutions (if it were good at this, people would not need eyeglasses in high-illumination scenarios, such as outdoors). Deconvolution is not magic. It uses the ability of the computer to perform deconvolutions to improve resolution at the expense of added noise. Therefore, the overall performance described by the metrics defined in this paper does not change as a result of a deconvolution operation [14]. Since most authors, including those of [30], measure MTF and efficiency anyway, all that is left is to measure veiling glare (or total transmission), and one can obtain full performance characterization based on physical parameters, without need for reference image based metrics such as SSIM. These proposed metrics (MTF and SNR) are in fact the industry standard for optical system characterization, and for good reason in our opinion [31,32].